\documentclass[aps,prl,twocolumn,showpacs,groupedaddress]{revtex4}

\usepackage{graphicx}
\usepackage{natbib}
\begin{document}

\title{Unified Brane Gravity:\\
Cosmological Dark Matter from Scale Dependent Newton Constant}

\author{Ilya Gurwich}
\email[Email: ]{gurwichphys@gmail.com}
\author{Aharon Davidson}
\email[Email: ]{davidson@bgu.ac.il}

\affiliation{Physics Department, Ben-Gurion University,
Beer-Sheva 84105, Israel}                 

\begin{abstract}
	We analyze, within the framework of unified brane gravity, the
	weak-field perturbations caused by the presence of matter on
	a 3-brane.
	Although deviating from the Randall-Sundrum approach, the
	masslessness of the graviton is still preserved.
	In particular, the four-dimensional Newton force law is recovered,
	but serendipitously, the corresponding Newton constant is shown
	to be necessarily lower than the one which governs FRW cosmology.
	This has the potential to puzzle out cosmological dark matter.
	A subsequent conjecture concerning galactic dark matter follows.
\end{abstract}

\pacs{
04.50.Kd,  % Higher-dimensional gravity and other theories of gravity
11.25.Db,  % Properties of perturbation theory
95.35.+d   % Dark matter
}
	
\maketitle    

%Section1
\subsection{\large{1. Introduction}}

The discovery of dark matter \cite{DMexp} continues to be a
challenging problem for astrophysics and cosmology.
Although many ideas from particle physics \cite{DMpart} have
been put forth, none of them so far have been able to provide a
convincing explanation for its mysterious nature and its
preponderance in the constitution of the Universe.
Besides, the search for a suitable particle candidate so far has
proved elusive.
A different approach proposes that dark matter is not matter at all
in the conventional sense but rather an artifact originating from a
deviation from general relativity.
Various forms of modified theories of gravity have been proposed
to explain the phenomenon \cite{DMtheory}.
However, all such theories are beset with their own problems.
In this context, it is important to note the remarkable coincidence in
the amounts of galactic and cosmological dark matter. 
This is natural for particle dark matter but has to be explained by
any full theory of artifact dark matter in modified theories of gravity.
Recently, the idea that brane world theories could provide an
explanation for the dark matter has been suggested \cite{DMbranes}.
Brane world theories have recently made great breakthroughs in
the area of reproducing some results of general relativity on the 
cosmic scale, as well as in deriving the low energy Newtonian limit
\cite{brane1,brane2,brane3,LBG}. 
The possibility that branes can naturally produce a solution to an
unsolved problem in gravity, such as dark matter, will generate a
great boost in the theory aside from being a significant achievement
and a good verification of the branes and extra-dimensions ideas.

In this paper we analyze the weak-field perturbations around a flat
background generated by matter on the brane.
We do so in the framework of unified brane gravity \cite{UBG},
following Dirac's prescription of careful variation in the region
of the brane \cite{Dirac}
(we present the basic principles in the next section).
After some remarks on the general scenario, we focus on the radial
case, thus studying the field created far from a source.
Our main results are as follows:
\begin{enumerate}
	\item We recover a Newtonian $1/r$ potential.
	\item The conventional (cosmological) Newton constant is suppressed
	by a constant factor greater than $1$. This difference between
	cosmological and radial Newton constants gives rise to natural
	cosmological dark matter.
	The amount of dark matter, characterized by the ratio between the
	two constants is an arbitrary parameter at this point.
\end{enumerate}
We later discuss the possibility of a transition scale between the
two Newton constants.
Such a transition will result in an effective deviation from the
Newtonian potential.
An observer who is unaware of this transition could interpret it
as a continuous distribution of dark matter.
To calculate the exact transition, one would need to analyze
weak-field perturbations around a cosmological brane, which is
very complicated.
We do calculate roughly the scale of the above transition without
the exact solution.
Remarkably the scale $\sim10^5$ ly (light years) is only 1 order
of magnitude above the experimental scale.
It remains to be seen whether this prediction is correct and whether
this transition will lead to flat rotation curves.

%Section2
\subsection{\large{2. The Basics of Unified Brane Gravity}}

Dirac has shown\cite{Dirac} that when performing variation of action
on a surface, around which one or more of the fields are discontinuous,
it is crucial to perform the variation in a coordinate system where
this surface remains static, to preserve the linearity of the variation.
Violating this principle results in nonlinear variation, and if this
problem is untreated, it would lead to incorrect equations of motion.
Dirac demonstrated this in his paper, where he performed both the
naive variation and the correct variation on a bubble model of an
electron, and showed that the naive variation results in a missing
term in the equations of motion.
In \cite{Karasik}, Karasik and Davidson demonstrated how the naive
variation would lead to a wrong Snell law, whereas the Dirac
prescription leads to the correct equation.

Unified brane gravity\cite{UBG} is based on the same action principle
as the standard brane models
(Randall-Sundrum, Dvali-Gabadadze-Porrati, and Collins-Holdom)
but following carefully Dirac's prescription for correct variation of
the brane.
We work in a coordinate system, where the variation of the bulk metric
on the surface of the brane has only 5 degrees of freedom,
\begin{equation}
	\delta g_{ab}=
	g_{AB,C}\delta y^{C}y^{A}_{,a}y^{B}_{,b}+
	2g_{AB}y^{A}_{,a}\delta y^{B}_{,b} ~,
\end{equation}
where $g_{AB}$ is the bulk metric and $y^{A}$ are the bulk coordinates.
This way, the brane remains undeformed during the variation.
The other degrees of freedom are not lost but are simply expressed by
the constraints that define the brane.
The equations of motion remain covariant and are independent of the
reference frame.
Using this principle of variation, the unified brane gravity field
equations for a $Z_2$ symmetric AdS bulk with an AdS scale $b^{-1}$
take the form
\begin{equation}
	\begin{array}{l}
	\displaystyle{\frac{1}{4\pi G_5}
	\left(K_{\mu\nu}-g_{\mu\nu}K\right)=}
	\vspace{4pt}\\
	\displaystyle{\frac{3b}{4\pi G_5}g_{\mu\nu}
	+\frac{1}{8\pi G_4}
	\left(R_{\mu\nu}-\frac{1}{2}
	g_{\mu\nu}R\right)+
	T_{\mu\nu}+\lambda_{\mu\nu}} ~.
	\end{array}
	\label{FieldEq2}
\end{equation}
In addition to the familiar terms (namely, the Israel junction term,
the brane surface tension, the Einstein tensor associated with the
scalar curvature ${\cal R}_4$, and the physical energy-momentum
tensor $T_{\mu\nu}=\delta {\cal L}_{matter}/\delta g^{\mu\nu}$
of the brane), unified brane gravity introduces $\lambda_{\mu\nu}$.
The latter consists of Lagrange multipliers associated with the
fundamental induced metric constraint
$g_{\mu\nu}(x)=g_{AB}(y(x))y^{A}_{,\mu}y^{B}_{,\nu}$.
In the above field equations, $\lambda_{\mu\nu}$ serves as a
geometric (embedding originated) contribution to the total
energy-momentum tensor of the brane.
If the variation would be performed naively, it would yield
\begin{equation}
	\lambda^{\mu\nu}=0~.
	\label{ubgwrong}
\end{equation}
This is reminiscent of Dirac's missing term.
A vanishing $\lambda_{\mu\nu}$ in Eq.(\ref{FieldEq2}) results in
the familiar Collins-Holdom equations
(Dvali-Gabadadze-Porrati with AdS bulk),
where these in turn contain the Randall-Sundrum and
Dvali-Gabadadze-Porrati equations as special cases.
Following Dirac's variation principle, we find that
$\lambda_{\mu\nu}$ is nonzero but is conserved, and its contraction
with the extrinsic curvature vanishes
\begin{equation}
	\lambda^{\mu\nu}_{~~;\nu}=0~, \quad
	\lambda_{\mu\nu}K^{\mu\nu}=0 ~.
	\label{ubgbasic2}
\end{equation}
This is reduced to the Regge-Tietelboim theory in the static bulk limit.
Since the Regge-Tietelboim theory was derived in a static bulk and is
the simplest of all brane models, it is very reassuring to have it as
a limit (the standard brane models are unsuccessful in that).

%Section3
\subsection{\large{3. General Perturbations and the Graviton }}

We begin with the simplest scenario of a four-dimensional flat brane of
positive tension embedded in five-dimensional AdS bulk
\begin{equation}
	\displaystyle{ds^{2}_{5}=dy^2+e^{-2b\left|y\right|}
	\eta_{\mu\nu}dx^{\mu}dx^{\nu}} ~.
\end{equation}
$b^{-1}=\sqrt{-6/\Lambda_5}$ denotes the AdS scale, $\eta_{\mu\nu}$
is the four-dimensional Minkowski metric, and the brane is conveniently
located at $y=0$.
Before turning to the main discussion concerning perturbations of this
brane, it is imperative to understand the full potential of the
unperturbed brane.
In the conventional Randall-Sundrum and Collins-Holdom scenarios, in
order to ensure its flatness, the brane has to be of positive
(or negative) tension:
\begin{equation}
	\sigma=\frac{3b}{4\pi G_5} ~.
	\label{tens}
\end{equation}
Unified brane gravity, although it requires the same, allows for one
more degree of freedom.

To see the point, first recall the unified brane gravity field
equations (\ref{FieldEq2},\ref{ubgbasic2}).
For a flat brane embedded in a five-dimensional AdS background,
which is the special case of interest, $K_{\mu\nu}=-b\eta_{\mu\nu}$.
In turn, Eq.(\ref{ubgbasic2}) simply implies that the corresponding
$\lambda_{\mu\nu}$ is traceless.
A traceless and conserved source serves as an effective (positive or
negative) radiation term.

The flatness of the unperturbed brane can be achieved the
conventional way, if the energy-momentum and the embedding terms
both vanish, that is, $T_{\mu\nu}=\lambda_{\mu\nu}=0$.
But now there exists the milder option $T_{\mu\nu}+\lambda_{\mu\nu}=0$.
Following the above, if (and only if) the real matter on the brane
exclusively consists of radiation, one can choose an appropriate
$\lambda_{\mu\nu}$ to cancel it out.
To be more specific, let our unperturbed flat brane host a constant
radiation density $\rho$, and choose the embedding counterterm to be
$\lambda^{0}_{\mu\nu}=-T^{0,rad}_{\mu\nu}=
-diag\left(\rho,\frac{1}{3}\rho, \frac{1}{3}\rho, \frac{1}{3}\rho\right)$.
This reflects the peculiarity that \emph{a flat brane can in fact be hot},
which is unique to unified brane gravity.
The perturbations around such a brane are expected to be quite different
from those around a Collins-Holdom brane, thus giving rise to new physics.
To study the perturbations induced by an arbitrary source
$\delta T_{\mu\nu}\equiv\tau_{\mu\nu}$, we find it useful to invoke
Gaussian normal coordinates, such that $\displaystyle{\delta g_{AB}=
h_{\mu\nu}}\delta^{\mu}_{A}\delta^{\nu}_{B}$ are the only allowed
nonzero components (and reserve the option of supplementing this gauge
later by the traceless nontransverse gauge).
It is important to keep in mind that, although $\tau_{\mu\nu}$ is
arbitrary, the perturbations of the metric are accompanied by
\textit{built-in} perturbations of all of the brane components that are
perturbed not by the source itself but rather by the shift in the
brane space-time structure.
For example, the radiation energy-momentum term must still satisfy the
conservation and traceless conditions, but it must be satisfied in the
new metric.
To this extent, the radiation term is corrected via a perturbation
$T^{rad}_{\mu\nu}=T^{0,rad}_{\mu\nu}+\delta T^{rad}_{\mu\nu}$ that
satisfies
\begin{eqnarray}
	\lefteqn{\partial^{\nu}\delta T^{rad}_{\mu\nu}=} \nonumber\\
	& & \eta^{\nu\lambda}
	\left(\Gamma^{\sigma}_{\lambda\mu}T^{0,rad}_{\sigma\nu}+
	\Gamma^{\sigma}_{\lambda\nu}T^{0,rad}_{\mu\sigma}\right)
	+h^{\nu\lambda}\partial_{\lambda}T^{0,rad}_{\mu\nu} , \\
	& & \eta^{\mu\nu}\delta T^{rad}_{\mu\nu} =
	h^{\mu\nu}T^{0,rad}_{\mu\nu},
\end{eqnarray}
to preserve conservation and tracelessness, respectively.
Here $\Gamma^{\lambda}_{\mu\nu}=
\frac{1}{2}\left(-\partial^{\lambda}h_{\mu\nu}+
\partial_{\mu}h^{\lambda}_{\nu}+
\partial_{\nu}h^{\lambda}_{\mu}\right)$ is the affine connection.
The above perturbation does not represent an addition of radiation
(which can be present independently via $\tau_{\mu\nu}$) but rather
a geometric effect.
By the same token, the embedding term is also perturbed via
$\lambda_{\mu\nu}=\lambda^{0}_{\mu\nu}+\delta\lambda_{\mu\nu}$
and satisfies
\begin{eqnarray}
	\lefteqn{\partial^{\nu}\delta \lambda_{\mu\nu}=} \nonumber\\
	& & \eta^{\nu\lambda}
	\left(\Gamma^{\sigma}_{\lambda\mu}\lambda^{0}_{\sigma\nu}+
	\Gamma^{\sigma}_{\lambda\nu}\lambda^{0}_{\mu\sigma}\right)
	+h^{\nu\lambda}\partial_{\lambda}\lambda^{0}_{\mu\nu} , \\
	& & \eta^{\mu\nu}\delta \lambda_{\mu\nu}=
	b^{-1}\delta K^{\mu\nu}\lambda^{0}_{\mu\nu},
\end{eqnarray}
However, since for a general perturbation $\delta K^{\mu\nu}$
is not proportional to $h^{\mu\nu}$, the term
\begin{equation}
	s_{\mu\nu}\equiv \lambda_{\mu\nu}+T^{rad}_{\mu\nu}
	=\delta\lambda_{\mu\nu}+\delta T^{rad}_{\mu\nu}
\end{equation}
is not necessarily zero.
One can furthermore verify that $s_{\mu\nu}$ is conserved
and not necessarily traceless:
\begin{equation}
	s\equiv \eta^{\mu\nu} s_{\mu\nu}=
	\frac{1}{2b}\lambda^{0}_{\mu\nu}\left(\frac{\partial}{\partial\left|y\right|}
	+2b\right)h^{\mu\nu} ~.
	\label{trace}
\end{equation}
The nonlocalized part of the perturbation equations is the same as the
familiar Randall-Sundrum case, since the bulk still follows the normal
five-dimensional Einstein equations
\begin{equation}
	\displaystyle{\left(\frac{\partial^2}{\partial\left|y\right|^2}
	-4b^2+e^{2b\left|y\right|}
	\fbox{\scriptsize 4}\right)
	h_{\mu\nu}=0,}
	\label{5dpert}
\end{equation}
where
$\fbox{\scriptsize 4}\equiv\eta^{\mu\nu}\partial_{\mu}\partial_{\nu}$
is the four-dimensional (unperturbed) d'Alembertian.
The localized part of the equation is
\begin{equation}
	\begin{array}{c}
	\displaystyle{\delta(y)\left[\frac{1}{8\pi G_5}
	\left(\frac{\partial}{\partial\left|y\right|}
	+2b\right)+\frac{1}{8\pi G_4}
	\fbox{\scriptsize 4}\right]h_{\mu\nu}}
	\vspace{4pt}\\
	\displaystyle{=\delta(y)\left(\tau_{\mu\nu}+
	s_{\mu\nu}\right).}
	\label{4dpert}
	\end{array}
\end{equation}
The propagation of modes into the bulk remains the same as in all
of the familiar cases.
Thus, we will be focusing on only the perturbations on the brane.
Performing separation of variables,
$h_{\mu\nu}=A(y)\bar{h}_{\mu\nu}\left(x^{\mu}\right)$
\footnote[1]{This is just a different parametrization to the mass
modes expansion.},
where we have normalized without loss of generality $A(0)=1$
and define $\displaystyle{\alpha=1+\frac{A^{\prime}(0)}{2b}}$.
Next let us separate the perturbation
$\bar{h}_{\mu\nu}=h^{(m)}_{\mu\nu}+h^{(u)}_{\mu\nu}$
to the standard term $h^{(m)}_{\mu\nu}$,
which follows the usual brane equation and thus admits the familiar
solutions and the new term $h^{(u)}_{\mu\nu}$, which is a direct
result of the additional effective source $s_{\mu\nu}$.
For $h^{(m)}_{\mu\nu}$, we can write
\begin{equation}
	\displaystyle{\left(\frac{\alpha b^2}{4\pi G_{RS}}
	+\frac{1}{8\pi G_4}
	\fbox{\scriptsize 4}\right)h^{(m)}_{\mu\nu}
	=\tau_{\mu\nu},}
	\label{0pert}
\end{equation}
where $G_{RS}=bG_5$ is the Randall-Sundrum gravitational constant
on the brane,
whereas for the new term
\begin{equation}
	\displaystyle{\left(\frac{\alpha b^2}{4\pi G_{RS}}
	+\frac{1}{8\pi G_4}
	\fbox{\scriptsize 4}\right)h^{(u)}_{\mu\nu}
	=s_{\mu\nu}.}
	\label{DGpert}
\end{equation}

Unfortunately, we cannot find a general Green function to
Eq.(\ref{DGpert}), because there is no closed-form expression of
$s_{\mu\nu}$ in terms of $h^{(u)}_{\mu\nu}$.
To that end, the only general prescription to solve Eq.(\ref{DGpert})
is perturbatively in $\rho$ (see Appendix A).
Despite not being able to find a general solution, we can get a clue
on its properties by taking the trace of Eq.(\ref{4dpert}) and
reorganizing the various terms
\begin{equation}
	\begin{array}{c}
	\displaystyle{\delta(y)\left(\frac{1}{8\pi G_5}
	\eta^{\mu\nu}-\frac{1}{2b}\lambda^{\mu\nu}_{0}\right)
	\left(\frac{\partial}{\partial\left|y\right|}
	+2b\right)h_{\mu\nu}}
	\vspace{4pt}\\
	\displaystyle{+\frac{1}{8\pi G_4}
	\eta^{\mu\nu}\fbox{\scriptsize 4}
	h_{\mu\nu}
	=\delta(y)\eta^{\mu\nu}\tau_{\mu\nu}.}
	\label{DGperttrace2}
	\end{array}
\end{equation}
Keeping in mind that $\lambda^{\mu\nu}_0\sim -\rho$ and $G_N\sim bG_5$,
by looking at the first term in the equation, one may expect that the
effective Newton constant may take the following form:
\begin{equation}
	\displaystyle{\frac{1}{G_N}=\frac{1}{G_{CH}}+
	\beta\frac{\rho}{b^2}~,}
	\label{GN0}
\end{equation}
where
\begin{equation}
	\displaystyle{\frac{1}{G_{CH}}=\frac{1}{G_{RS}}+\frac{1}{G_4}}
	\label{CHNewtonconst}
\end{equation}
is the effective Newton constant in the Collins-Holdom scenario and
$\beta$ is a dimensionless constant.
In the next section, we show that this prediction is indeed true and,
interestingly, $\beta$ is geometry dependent.

Although we did not obtain a propagator for the graviton, the form of
the equation looks all too similar to the usual brane equations and
along with Eq.(\ref{DGperttrace2}) suggests that, despite deviating from
the standard Randall-Sundrum scenario, the graviton propagator remains
the same.
In the following section, we show that the Newtonian potential is recovered
for large $r$ and thus prove that the graviton's zero-mode is massless.

%Section4
\subsection{\large{4. Static Radial Source}}

In all studies of gravitational perturbations, the pointlike radial
source is of special interest.
Since an exact radial solution is missing in all brane theories, the best
idea we have for a radial potential comes from perturbative treatment.
We solve the equations far from the source, in the region where
$\tau_{\mu\nu}=0$.

For the radial case, we show that an exact (non perturbative in $\rho$)
weak-field solution can be obtained.
This is mainly due to the fact that we are able to express
$s_{\mu\nu}$ explicitly.
We choose to work in a traceless Gaussian frame.
For a radially symmetric perturbation, we can choose, in addition to the
Gaussian traceless gauge, the radial gauge
\footnote[2]{ Since any static radial perturbation can be brought
to the form of $ds^2=-f(r)dt^2+g(r)dr^2+r^2d\Omega^2$, we can impose
$h_{\theta\theta}=h_{\phi\phi}=0$.}.
It follows that $s_{\theta\theta}=s_{\varphi\varphi}=0$.
Solving the conservation equation for $s_{\mu\nu}$ along with
Eq.(\ref{trace}) and gauging following the above, we have
\begin{equation}
	\displaystyle{s_{tt}(r)=s_{rr}(r)=
	-\frac{1}{4}s(r)+\frac{1}{2r^2}\int drrs(r).}
	\label{DGsource}
\end{equation}
From Eq.(\ref{DGpert}), we see that since $s_{tt}=s_{rr}$ it follows that
$h^{(u)}_{tt}=h^{(u)}_{rr}\equiv h^{(u)}$.
This is the familiar form of radial fluctuations.
Finally, $h^{(m)}_{\mu\nu}$ constitutes the familiar Collins-Holdom solution.
The exact solution is quite complicated, but to first order in $1/r$, the
solution simply yields
\begin{equation}
	\displaystyle{h^{(m)}_{tt}=h^{(m)}_{rr}\cong\frac{2G_{CH}M}{r},}
	\label{CHpert}
\end{equation}
where $M=\int d^3x\tau_{tt}$ is the mass of the source.
Substituting Eq.(\ref{trace},\ref{DGsource},\ref{CHpert}) along with
\begin{equation}
	\fbox{\scriptsize 4}=\frac{1}{r^2}\frac{d}{dr}
\left(r^2\frac{d}{dr}\right)
\end{equation}
into Eq.(\ref{DGpert}),
we can write the equation for $h^{(u)}$
\begin{equation}
	\begin{array}{c}
	\displaystyle{\kappa^2_4rh^{(u)\prime\prime\prime}+
	4\kappa^2_4h^{(u)\prime\prime}}
	\vspace{4pt}\\
	\displaystyle{+
	\left(\frac{2\kappa^2_4}{r}+\left(k-\frac{2}{3}
	\alpha\rho\right)r\right)
	h^{(u)\prime}}
	\vspace{4pt}\\
	\displaystyle{+2kh^{(u)}=
	-\frac{4G_{CH}M\alpha\rho}{3r},}
	\end{array}
	\label{radpert}
\end{equation}
where $\displaystyle{\kappa^2_4\equiv\frac{3}{16\pi G_4}}$ and
$\displaystyle{k\equiv\frac{\alpha b^2}{2\pi G_{RS}}}$. 
The solution of physical relevance is the nonhomogeneous one, namely,
\begin{equation}
	h^{(u)}=
	-\frac{2G_{CH}M}{\displaystyle{1+\frac{3b^2}{4\pi G_{RS}\rho}}}\frac{1}{r}.
\end{equation}
The full perturbation
$\bar{h}_{\mu\nu}=h^{(m)}_{\mu\nu}+h^{(u)}_{\mu\nu}$ is therefore
\begin{equation}
	\bar{h}_{tt}=\bar{h}_{rr}=
	\frac{1}{\displaystyle{1+\frac{4\pi G_{RS}\rho}{3b^2}}}\frac{2G_{CH}M}{r}.
\end{equation}
It is important to note that it is only due to the solution
being independent of $\alpha$ that we can proceed without integrating
over all the mass modes.
The Newtonian potential is thus recovered, giving us further reassurance
that the graviton is indeed massless, since a mass term in the propagator
would have generated an exponential decay.
The associated Newton constant is
\begin{equation}
	G^{r}_N=\frac{G_{CH}}{\displaystyle{1+\frac{4\pi G_{RS}\rho}{3b^2}}} ~,
	\label{GN1}
\end{equation}
where the $r$ index stands for radial.

Now that the mathematics has been understood, we return to physics.
Alone, Eq.(\ref{GN1}) has nothing new to offer.
However, gravitational measurements in our Universe, although they began
with the Solar System, which is physically a radially symmetric system,
are now quite based in the field of cosmology as well.
We recall (see Appendix B) the cosmological result for expansion around
a flat background gives an FRW solution with an associated Newton constant
\begin{equation}
	\displaystyle{\frac{1}{G^{c}_N}=\frac{1}{G_{RS}}+\frac{1}{G_4}+
	\frac{4\pi\rho}{3b^2},}
	\label{GN3}
\end{equation}
where the $c$ index stands for cosmological and $\rho$ here has the exact
same role of background radiation.
Equation (\ref{GN1}) can also be written as
\begin{equation}
	\displaystyle{\frac{1}{G^{r}_N}=\frac{1}{G_{RS}}+\frac{1}{G_4}+
	\frac{4\pi\rho}{3b^2}\left(1+\frac{G_{RS}}{G_4}\right).}
	\label{GN2}
\end{equation}
Now, if we further assume that the role of radiation in our case is also
played by the background radiation from cosmology, we can compare the two
results.
First of all, since we do have bounds on $b$ from both particle and
gravitational localization, we can clearly state that the term
$\displaystyle{\frac{\rho}{b^2}}$ is negligible in both equations.
This means that $G^{c}_{N}=G_{CH}$, whereas
\begin{equation}
  \displaystyle{\frac{1}{G^{r}_N}=\frac{1}{G^{c}_N}+
	\frac{4\pi\rho}{3b^2}\frac{G_{RS}}{G_4}.}
	\label{GN4}
\end{equation}
The last term in the radial gravitational constant would have been
negligible if not for the factor $\displaystyle{\frac{G_{RS}}{G_4}}$.
We have no experimental or theoretical bounds on the latter ratio.
In fact, the proposed self-accelerated Dvali-Gabadadze-Porrati solution
for the cosmological constant requires this quantity to be very large.
If it is large enough, then the above term can be significant in the
calculation of the Newton constant.
Thus, in principle, we have a real difference between the cosmological
and the radial gravitational constants,
the radial constant being necessarily lower.
However, historically, the Newton constant was measured in radial systems
(the Solar System).
Thus \emph{an observer that is unfamiliar with this physics would
interpret this effective growth of the gravitational constant as missing
cosmological mass} (since, in general relativity, mass is inseparable from
the gravitational constant),
thus bringing him to the phenomenon of cosmological dark matter,
\emph{without facing dark matter in the Solar System}.

Although we have not shown it here
(this is a conjecture subject to future research),
when solving the perturbation equations around a cosmological background,
one expects the two branches of the solution, one being the $G^{r}_{N}$
and the other $G^{c}_{N}$, to be connected, creating some sort of transition
between them.
Such a transition, to an observer that is unaware of this effect, will seem as
a gradual increase of mass, that may result in flat rotation curves.
Although the exact solution to fluctuations around a cosmological brane is
highly complex, we can give a rough estimate to the typical scale of such a
transition.
We assume the scale to be roughly in the region where the cosmological and
radial curvatures are of the same order of magnitude, so that the
cosmological and radial solutions \emph{"mix"}.
The radial curvature is of the order $\displaystyle{\frac{r_s}{r^3}}$,
$r_s$ being the Schwarzschild radius and the cosmological curvature is of
the order of $H^2$, $H$ being the Hubble constant.
The scale of the predicted transition is therefore
\begin{equation}
	\displaystyle{r_{dm}\sim\left(r_st^{2}_{Hubble}\right)^{1/3}\propto M^{1/3},}
	\label{FRC}
\end{equation}
where $t_{Hubble}$ is the age of the Universe.
When this scale is calculated for the Sun, the result is 100 ly, which is
way beyond the scale of the Solar System.
At these distances, other stars contribute, and thus the effect is
unmeasurable today.
For a galactic mass, on the other hand, the result is of the order of
$10^5$ ly, which is only 1 order of magnitude higher than the real
galactic scale.
One needs to remember that it is only a rough estimate and also that
galaxies are not radial systems and are composed of many stars,
each giving an effect on the scale of about 100 ly, so that the combined
effect may be closer than the above result, to give the exact scale of
flat rotation curves.

%Section5
\subsection{\large{5. Summary and Conclusions}}

We have studied the behavior of weak-field perturbations around a flat
brane, in the framework of unified brane gravity.
It was shown that, even for the most general perturbation, the novel
embedding term is "harmless" and the graviton propagator is intact,
leaving the graviton massless.
We verify this result, in particular, for a spherically symmetric source,
where the conventional Newtonian potential $1/r$ is recovered.
However, upon a closer examination, we see that, although the functional
form of the potential is standard, the gravitational constant differs
from the one found in cosmology.
Furthermore, the radial gravitational constant is necessarily lower than
the cosmological one.
For an observer, familiar only with Einstein's general relativity, this
would be immediately interpreted as \textbf{cosmological dark matter}.
This can also be the source of galactic dark matter.
The flat rotation curves may simply represent the transition between the
radial and cosmological gravitational constants.
The scale of the suggested flat rotation curves is predicted in this case
to be of the order of $\left(r_st^{2}_{age}\right)^{1/3}$.
When evaluated for a galactic mass, this is indeed close to the galactic
scale.
Despite this transition being the natural outcome of the two different
gravitational constants, there is no reason why such a transition would
generate flat (rather than some general form) rotation curves, and the
flatness of the rotation curves is wishful thinking at this point.

Although the radial dark matter solution is completely speculative in
this paper, the cosmological dark matter is fully postulated.
The only thing that is arbitrary is the amount of dark matter.
This is due to the arbitrariness of $\displaystyle{\frac{G_{RS}}{G_4}}$.
In fact, in order to account for the right amount of dark matter, we
would need an extremely large $G_5$, implying a very low five-dimensional
plank mass $M_5\approx\left(\rho l\right)^{1/3}$.

\appendix

\phantomsection \addcontentsline{toc}{part}{Appendix}

%section Appendix A
\subsection{\large{Appendix A: Perturbative Method}}

We can expand the solution to Eq.(\ref{DGpert}) via
\begin{equation}
	\displaystyle{h^{(u)}_{\mu\nu}=
	\sum^{\infty}_{i=1}h^{(i)}_{\mu\nu}}
	\label{series}
\end{equation}
and
\begin{equation}
	\displaystyle{\left(\frac{\alpha b^2}{4\pi G_{RS}}
	+\frac{1}{8\pi G_4}
	\fbox{\scriptsize 4}\right)h^{(i)}_{\mu\nu}
	=s^{(i)}_{\mu\nu},}
	\label{DGpert_series}
\end{equation}
where, for $i>1$,
\begin{equation}
	\displaystyle{s^{(i)}\equiv s^{(i)}_{\mu\nu}\eta^{\mu\nu}=
	\alpha\lambda^{\mu\nu}_{0}h^{(i-1)}_{\mu\nu}}
	\label{trace_seriesi}
\end{equation}
and
\begin{equation}
	\displaystyle{s^{(1)}\equiv s^{(1)}_{\mu\nu}\eta^{\mu\nu}=
	\alpha\lambda^{\mu\nu}_{0}h^{(m)}_{\mu\nu}.}
	\label{trace_series1}
\end{equation}

%section Appendix B
\medskip\subsection{\large{Appendix B: Cosmological Gravitational Constant}}

In \cite{UBG} we have proven that, when expanding the cosmological
equations around a flat background with positive tension and radiation
density of Eq.(76),
\begin{equation}
	\displaystyle{\rho (a)=
	\sqrt{\frac{2}{-\Lambda_{5}}}
	\frac{\omega}{a^{4}} ~,}
	\label{rhoflat}
\end{equation}
where $\omega$ is a constant.
The resulting FRW equation was given by Eq.(81):
\begin{equation}
	\displaystyle{\widetilde{\rho} =
	\left(\frac{1}{8\pi G_{4}}+\sqrt{\frac{6}{-\Lambda_{5}}}
	\left(\frac{1}{8\pi G_{5}}+
	\frac{\rho}{6b}\right)
	\right)\epsilon~,}
\end{equation}
where $\displaystyle{\epsilon=3\frac{\dot{a}^{2}+k}{a^{2}}}$
and, therefore,
\begin{equation}
	\displaystyle{\frac{1}{G^{c}_N} =
	\frac{1}{G_{4}}+\frac{1}{G_{RS}}
	+\sqrt{\frac{2}{-\Lambda_{5}}}\frac{8\pi\omega}{\Lambda_{5}a^{4}}
	~.}
	\label{GN5}
\end{equation}
We would like to express the last term in Eq.(\ref{GN5}) in terms
of $\rho$ and $b$ and, therefore,
\begin{equation}
	\displaystyle{\frac{1}{G^{c}_N}=\frac{1}{G_{CH}}+
	\frac{4\pi\rho}{3b^2}.}
	\label{GN6}
\end{equation}

\acknowledgments{}
The authors thank Professors Philip Mannheim, Eduardo Guendelman
and especially our colleague Shimon Rubin for enlightening discussions and constructive
comments. A special thanks to Wali Kameshwar for constructive comments that have
significantly improved this paper.

\end{document}